\documentclass[aps,amsmath,amssymb,notitlepage]{revtex4-1}
\usepackage{bm}% bold math
\usepackage[english]{babel}
\usepackage{graphicx}
\usepackage{graphics}
\usepackage{slashed}
\usepackage{placeins}
\graphicspath{{FIGURE/}}
\newcommand{\bb}[1]{\boldsymbol{#1}}

\begin{document}
 
\hfill FTUV-13-0109 

\hfill IFIC/13-62

\title{The half-skyrmion phase in a chiral-quark model}
\author{Valentina Mantovani Sarti$^{1}$ and Vicente Vento$^{2}$ }
\affiliation{$(1)$~Dipartimento di Fisica, Universita’ di Ferrara and INFN, Sez. Ferrara, 44100 Ferrara, Italy. \\$(2)$~Departamento de F\'{\i}sica Te\'orica
and IFIC, Universidad de Valencia-CSIC, 46100 Burjassot (Valencia), Spain}

\begin{abstract}
The Chiral Dilaton Model, where baryons arise as non-topological solitons built from the interaction of quarks and chiral mesons, shows  in the high density low temperature regime a two phase scenario in the nuclear matter phase diagram. Dense soliton matter  described by the Wigner-Seitz approximation generates a periodic potential in terms of the sigma and pion fields that leads to the formation of a band structure. The analysis up to three times nuclear matter density shows that soliton matter undergoes two separate phase transitions: a delocalization of the baryon number density leading to $B=1/2$ structures, as in skyrmion matter, at moderate densities, and quark deconfinement at larger densities. This description fits well  into the so-called quarkyonic phase  where, before deconfinement, nuclear matter should undergo structural changes involving the restoration of fundamental symmetries of QCD. 
\end{abstract}

\maketitle

\section{Introduction} 

The availability of accurate and precise experimental data from heavy-ion collisions at relativistic and ultra-relativistic energies
permits to gain more and more insight into the theoretical description of nuclear matter under extreme conditions.
It has been suggested that during these very high energy collisions, ordinary hadronic matter undergoes a phase transition, or in general a 
"significant change", that leads to the formation of the so-called \textit{Quark-Gluon-Plasma}~\cite{Heinz:2000bk}.

\begin{figure}[b]
\center
\includegraphics[width=0.4\textwidth]{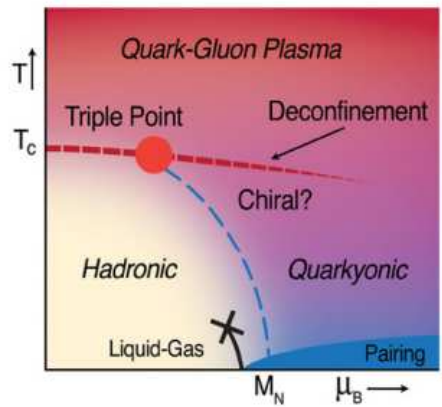} 
\caption{The phase diagram of strongly interacting matter, taken from~\cite{Andronic:2009gj}}\label{PhaseDiag}
\end{figure}

At zero or at least at very low baryon densities, we expect a smooth transition
from a gas of confined hadrons to a plasma of free gluons and quarks (see the left side of Fig.~\ref{PhaseDiag}). Experimentally this area is covered by the LHC
collider and it has also been investigated by RHIC in the past years.
From the theoretical point of view, the description of such a transition  at very high energies involves both  perturbative and non-perturbative aspects of the exact
underlying theory, \textit{Quantum Chromodynamics} (QCD). 
Lattice QCD, unfortunately only at very low densities, provides precise information concerning the deconfinement 
transition from hadrons to quarks and gluons~\cite{Cheng:2006aj, Cheng:2006qk, Borsanyi:2010cj}.
Moreover, perturbative techniques~\cite{Blaizot:2001nr, Lacour:2009ej} 
are able to provide estimates on the properties of matter formed above the critical temperature $T_c$.

Moving along the $\mu_B$-axis of the phase diagram, we enter the area of higher baryon densities where 
the non-perturbative features of QCD are dominant. At the moment little is known from experiments in heavy-ion collisions and astrophysics, about the behaviour of QCD at these large densities.
The gap between the underlying theory and the phenomenology in this regime is 
occupied by effective theories, whose Lagrangians contain the relevant degrees of freedom at that energy scale implementing the fundamental symmetries of QCD. Moreover, the intrinsic numerical difficulties in the evaluation of observables  non-perturbatively leads to the lack of any reliable expansion scheme and one is forced to
resort to model Lagrangians~\cite{Nambu:1961tp, Skyrme:1961vq, Diakonov:1987ty}.
As a result the high baryon density region allows for  various scenarios associated with the
restoration of symmetries such as chiral symmetry or scale invariance.
For example, it has been suggested, using the limit of large number of colours $N_c$ ~\cite{L.McLerran2007}, that before
deconfinement, the so called quarkyonic transition will take place, where
several  symmetries are eventually restored leading to a rich set of phases.

One of the most successful theories rooted in large $N_c$ is the Skyrme model~\cite{Skyrme:1961vq, Skyrme:1962vh}.
Even though the model does not include quarks and gluons but only pions, it has been widely used to study dense matter and in particular
the chiral symmetry restoration at large densities, which should occur not very far from the deconfinement.
The ground state matter built upon this Lagrangian at low density is described by a crystal of localized single-skyrmions, as shown 
by many authors~\cite{Klebanov:1985qi,  Jackson:1987sx, Goldhaber:1987pb, Walhout:1987ce}. 

As the density increases the  crystal lattice of skyrmions undergoes a phase transition leading to another crystal structure made of  well localized skyrmions but carrying only one half of the baryon number ~\cite{Castillejo:1989hq, Kugler:1989uc}. To be more precise, imagine our single skyrmions, each one centered on the vertices of a cubic lattice (of size $L$) and the corresponding baryon density profiles peaking at the same points and going to zero on the links connecting the skyrmions. 
As the lattice size decreases (the density, $\rho_B=L^{-1/3}$, increases) the baryon density starts to develop a new local maximum at the center of the joining links, leading to a delocalization of the baryon charge and to the so called \textit{half-skyrmion} phase. The Skyrme Lagrangian has been applied to study in detail the chiral phase
transition at finite density~\cite{Park:2002ie, Lee:2003aq, Park:2003sd, Park:2008xm, Park:2009bb} with the implementation of the scale invariance in the model by
introducing a new scalar field, the dilaton~\cite{Lee:2003eg, Park:2008zg}. In these works it has been shown that in the high density phase, the half-skyrmion phase, 
described by cubic-centered half-skyrmion crystal, the expectation value of the $\sigma$ field drops to zero signalling the restoration of chiral symmetry.

On the basis of these results obtained in a model with only meson fields, we decided to understand whether the same delocalization mechanism, signalling a phase transition, takes place in the Chiral Dilaton Model (CDM) where baryons are instead made of quarks interacting with sigma and pion fields. 
The model we consider has been already studied in detail in~\cite{Drago:2011hq} and the corresponding simplified density Lagrangian~\cite{E.K.Heide1994,G.W.Carter1998,G.W.Carter1997,G.W.Carter1996,Bonanno:2008tt} reads,
\begin{align}\label{lagrCDMchiral}
\mathcal{L}  = \bar{\psi}  [i \gamma ^\mu\partial _{\mu}-
g_\pi(\sigma + i \boldsymbol{\pi}\cdot \boldsymbol{\tau}\gamma _{5})]\psi 
  +\dfrac{1}{2}(
\partial_{\mu}\sigma \partial^{\mu}\sigma + \partial_{\mu}\boldsymbol{\pi}\cdot \partial^{\mu}\boldsymbol{\pi})
-V(\sigma,\mathbf{\pi}).
\end{align}
The potential is given by,
\begin{align} \label{potfro}
& V(\sigma,\bb{\pi}) =  \lambda_1 ^2 (\sigma^2 +\boldsymbol{\pi}^2)-\lambda_2 ^2\,\ln(\sigma^2 + 
\boldsymbol{\pi}^2)-\sigma_0 m_\pi ^2 \sigma,
\end{align}
where,
\begin{align}
& \lambda_1 ^2=\dfrac{1}{2}\dfrac{B \delta \phi_0 ^4 + m_\pi ^2 \sigma_0 ^2}{\sigma_0 ^2}=\dfrac{1}{4}(m_\sigma ^2+m_\pi ^2), \nonumber\\
& \lambda_2 ^2=\dfrac{1}{2}B \delta \phi_0 ^4= \dfrac{\sigma_0 ^2}{4}(m_\sigma ^2-m_\pi ^2). \nonumber
\end{align}

Here $\sigma$ is the scalar-isoscalar field, $\bb{\pi}$ is the pseudoscalar-isotriplet meson field, $\phi$ is
the dilaton field  and $\psi$ describes the isodoublet quark fields. This Lagrangian density, besides being invariant under  chiral symmetry, is also invariant under another fundamental symmetry in QCD, scale invariance, which is  spontaneously broken,  and has been implemented above following Refs.~\cite{J.Schechter1980,Migdal:1982jp,E.K.Heide1992}.

The dilaton field will be frozen in the present calculation at its vacuum value $\phi _0$. This approximation  is based on the results obtained in~\cite{Bonanno:2008tt, G.W.Carter1998}, which showed that at low temperatures the dilaton remains close to its vacuum value even at large densities. 

The constants B and $\phi_0$ in the potential are fixed by choosing a value
for the mass of the glueball and for the vacuum energy,
while $\delta = 4/33$ is given by the QCD beta function and it
represents the relative weight of the fermionic and
gluonic degrees of freedom. The vacuum state is chosen, as usually, at $\sigma_0=f_\pi$ and $\bb{\pi}=0$.

\section{The Chiral Dilaton Model at finite density}  

Recently, it has been shown that the  CDM with hadronic degrees of freedom provides a 
good description of nuclear physics at densities around $\rho_0$ and   the gradual restoration of chiral
symmetry at higher densities~\cite{Bonanno:2008tt}.  In~\cite{Drago:2011hq} the authors used this model interpreting 
the fermions not as nucleons but as quarks and they
showed that, at finite density, the new potential allows to reach densities higher than the ones obtained in the linear $\sigma$-model.

In order to study the CDM at finite density we use the so-called \textit{Wigner-Seitz approximation}~\cite{Wigner:1933zz}, coming 
from solid state physics, which has already been applied to  non-topological soliton models~\cite{Birse:1987pb, Reinhardt:1985nq, U.Weber1998}, 
chiral soliton models~\cite{Glendenning:1986fy, Hahn:1987xr, U.Weber1998, P.Amore2000} and 
also to the Skyrme model~\cite{Wuest:1987rc, Amore:1998fp}.
This approximation consists in replacing the cubic lattice (as the one used in the Skyrme model) by
a spherical symmetric one where each baryon is centered on a spherical cell of radius $R$. The finite density effects are provided by the
requirement of specific boundary conditions on the fields at the surface of the sphere.
In the literature  different sets of boundary conditions are used  depending on which symmetry of the crystal is imposed~\cite{U.Weber1998, P.Amore2000}. In our work we adopted the choice of Ref.~\cite{U.Weber1998}, which relates the boundary conditions to the parity operation, $\bb{r}\rightarrow -\bb{r}$. In this scheme the odd fields, such as the upper Dirac component and the pion,  are forced to vanish at $R$, while in the case of the even fields (upper Dirac component and sigma field)  their first derivative is forced to vanish at the boundary. A detailed description of the finite density results achieved with the CDM, can be found in Ref.~\cite{Drago:2011hq}.

The periodic potential generated by the meson field configurations in which the quarks move leads to the formation of a band structure, with energy bands and gaps, as happens for nearly free electrons moving in a crystal structure~\cite{kittel1986introduction}. In solid state physics the filling of these bands determines the distinct behaviour of insulators, metals and semiconductors.
For periodic potentials, the spinor eigenfunction of the Schroedinger equation must satisfy the Bloch theorem:
\begin{equation}
\psi _{\boldsymbol{k}} (r)=e^{i \boldsymbol{k}\cdot \boldsymbol{r}}\Phi_{ \boldsymbol{k}}(r),
\end{equation}
where $\boldsymbol{k}$ is the crystal momentum (which for the ground state
is equal to zero) and $\Phi_{ \boldsymbol{k}}(r)$ is a spinor that has
the same periodicity of the lattice.

The definition of the band width and the filling of the band are non-trivial ~\cite{U.Weber1998,Barnea:2000nu}.
We adopt a simple procedure, following Ref.~\cite{Birse:1987pb},  which obtains the band width 
by imposing that the lower Dirac component vanishes at the boundary. In Ref.~\cite{U.Weber1998} it is shown that
the eigenvalue corresponding to the top of the band, $\epsilon_{top}$, obtained within this approximation is an
upper limit to the top of the band and that the true top lies about halfway between this upper limit and the bottom.

Concerning the filling of the band, when working with chiral solitons
at mean field level, the relevant quantum number is the grand-spin $G$
and the lower band corresponds to $G=0$. The only degeneracy remaining
is color and therefore three quarks per soliton  completely
fill the band. Comparing with  the electron case, since the band is totally
filled, our soliton lattice acts as a color insulator and the quarks are well localized in each cell. 
In the next section we describe how this picture changes when we increase the density.

\section{The Chiral Dilaton Model at high density}

 The phase diagram for the Skyrme model at moderate temperatures and large densities exhibits a two phase scenario described by a crystal of well localized skyrmions at low density and a crystal of half-skyrmions at high density in which chiral symmetry is restored. \textit{Is there a similar scenario in a model with quarks?}

\begin{figure}[b]
\center
\includegraphics[width=0.4\textwidth]{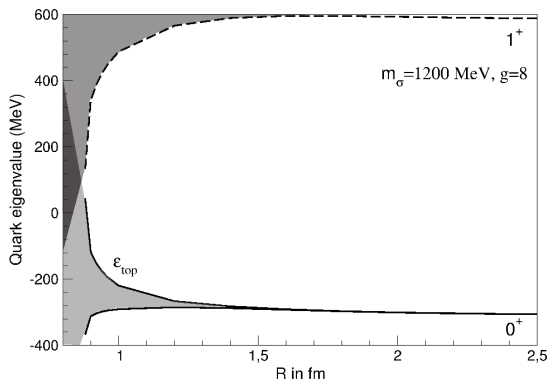} 
\caption{Quark eigenvalue as a function of the cell radius $R$ in the CDM. The shaded area represents the band estimated following Ref.~\cite{Birse:1987pb}. The first excited state $1^+$ and the corresponding lower part of the band are also shown.}\label{deconf}
\end{figure}

In the skyrmion description, as the system reaches high densities, a delocalization of baryon number takes place. We can
imagine that this mechanism translates into a sort of modification of the baryon density profile in the CDM.
In Fig.~\ref{deconf} we plot the quark eigenvalue for the CDM as a function of the cell radius $R$.
The line labeled $\varepsilon_{top}$ corresponds to the estimate of the top of the band proposed by Birse in Ref.~\cite{Birse:1987pb}, while the dashed line represents the eigenvalue for the first excited state.
Proceeding from lower densities, namely larger values of $R$, the system behaves as a lattice of
 well localized solitons with quarks and meson fields confined in each cell.
As the radius of the cell shrinks, the band gets wider and quarks from neighbouring sites, previously enclosed 
in their cell, become free to move along the lattice. Moreover, quarks also are able to move between the lower band and the excited band. This happens in a range of cell radii going from approximately  $0.8- 1.2$ fm, corresponding 
to a baryon density of roughly $\rho_B/\rho_0\approx0.85-3$, where $\rho_0$ is nuclear saturation density. 

If we look with more detail, the figure shows two distinct phenomena, both correlated to the presence of a band structure. On one hand
the exchange of quarks between the lower band $G=0^+$ and the first excited one $G=1^+$ occurring at densities larger than
$3\rho_0$ can be interpreted as a signal of \textit{deconfinement}, since the crystal becomes a color conductor ~\cite{Hahn:1987xr}.
On the other hand, the plunging  of the excited state into the lower band and the sharing between neighbouring cells, as will be shown next, is
the origin of a \textit{delocalization} of the baryon number carried here only by quarks, leading to an apparent $B=1/2$ phase. A word of caution, this
phase should  not be confused with the half-skyrmion phase describing a rigid crystal structure \cite{ Park:2009bb}, although we suspect that 
the delocalization mechanism is the seed for the quarkyonic interpretation of the half-skyrmion phase.

\begin{figure}[t]
\center
\fbox{\includegraphics[width=0.9\textwidth]{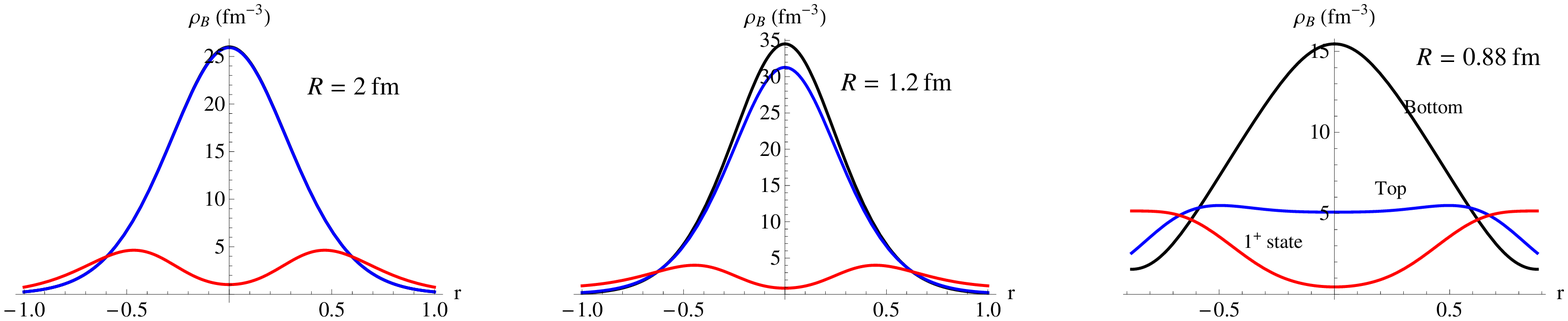}} 
\caption{Baryon density profiles inside the Wigner-Seitz cell, for several values of $R$. The profiles are shown for the ground state $0^+$ (black line), the top of the lower band (blue line) and the first excited state $G= 1^+$ (red line). The three panel are chosen in order to
show, going from the left to right, the confined $B=1$ phase ($R\geq 2$ fm), the delocalized $B=1/2$ phase ($R\leq 1.2$ fm) and finally the deconfined phase ($R\leq 0.9$ fm).}\label{deloc}
\end{figure}

The modification of the baryon density profile is shown in Fig.~\ref{deloc}. We plot, for several values of the cell radius $R$, 
the baryon density profile inside one cell in three cases: a $G=0^+$ state, a state at the top of the band and an excited  $G=1^+$ state.

\begin{figure}[b]
\center
\fbox{\includegraphics[width=0.2\textwidth]{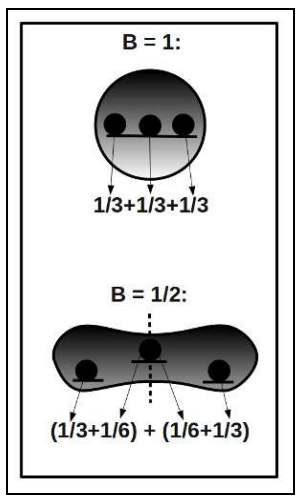}} 
\caption{Representation of the delocalization process. In the upper part, the quarks in the $B=1$ soliton lie in the ground state $0^+$ and the baryon density presents a maximum at the center of the cell (upper panel in Fig.~\ref{deloc}). As the radius shrinks, some quarks may jump into the first excited level $1^+$, leading to a delocalization of baryon number as shown in the lower part of the figure.  At the same time, as the radius decreases, the band gets wider and quarks are shared between neighbouring cells. The interplay between these two mechanisms leads to the $B=1/2$ phase.}\label{balls}
\end{figure}

At very low densities (for $R= 2$ fm, $\rho_B\approx 0.2\rho_0$) the solitons are well localized inside the cell, all the quarks occupy the lower state, as can be seen in the left panel of Fig.~\ref{deloc}. The baryon density at  the top of the band almost coincides with that at the bottom, and is clearly peaked at the center of the cell  vanishing at the edges. As the radius shrinks, the band gets wider (middle panel of Fig.~\ref{deloc}) as the $G=1^+$ state  plunges into the lower band (see Fig.~\ref{deconf}). Now, the quarks are free to move to the excited state whose baryon density profile shows a valley at the center of the cell and two bumps close to the edges. We can imagine a $B=1$ soliton with one quark in the $G=1^+$ state and the other two lying in the lower band. This mechanism provides a delocalization of the baryon number from the center of the cell to the edges and leads to a $B=1/2$ scenario as shown in Fig.~\ref{balls}. In our calculation, scale invariance cannot be restored due to the frozen dilaton's dynamics, and therefore chiral symmetry restoration cannot take place. 

As we reach higher densities, a new scenario takes place,  as shown in the right panel in Fig.~\ref{deloc}, the band broadens and the baryon densities for  both the $G=0^+$ band and the top of the band do not vanish anymore at the boundaries of the cell, allowing a continuous flow of quarks between all cells. This scenario represents the deconfinement of quarks, since they are now free to migrate everywhere in the lattice. This happens before the breakdown of the solution, which  indicates that beyond a critical density the field equations of the model only admits a  trivial solution, i.e.  plane waves propagating in the crystal~\cite{Barnea:2000nu}.
As clearly stated in Ref.~\cite{Barnea:2000nu}, the loss of non-trivial solutions at large densities is associated with the Wigner-Seitz approximation. 

The Wigner-Seitz approximation, although very useful and widely used, shows also limitations compared to the skyrmion lattice, since it does not take into account the long-correlation between neighbouring baryons coming from the presence of pions and does not allows the inclusion of different isospin configurations.

\section{Conclusions and Outlook} 

We used a Lagrangian with quarks degrees of freedom and meson fields based on chiral and scale invariance to analyse the
possible presence of a $B=1/2$ phase at large baryon densities, similarly to what happens in skyrmionic matter~\cite{Castillejo:1989hq, Kugler:1989uc, Lee:2003eg, Park:2008zg}. In the Skyrme model, since the baryon number is  carried by the skyrmion, the half-skyrmion phase is
strictly connected to the topological modification of the pion fields inside the lattice, as the system reaches high densities. In this work, motivated by the so-called quarkyonic phase~\cite{L.McLerran2007},  we have shown that a similar mechanism occurs in a model  like the CDM , which contains explicit fermionic degrees of freedom. In our case the baryon number is carried only by quarks and the $B=1/2$  or half-skyrmion phase is achieved by a delocalization of the baryonic charge, i.e by a modification of the baryon density profile, as we move to high densities.

To describe the CDM at large densities we have adopted the Wigner-Seitz approximation, in which solitonic matter is described by a lattice of spherical cells of radius $R$ and finite density effects are obtained by imposing boundary conditions on fields at the boundaries of the cells.
The presence of a periodical potential generated by the sigma and the pion fields leads to a band structure, in which quarks move.
As the cell radius decreases,  the baryon density increases, we have shown that the development of the quark band gives rise to two separate phenomena.
As the system approaches densities around $\rho_0$, the band starts to get wider and the quarks, still in the lowest state, are free to move inside the cell. As the quarks get squeezed inside the cell, they get excited and as the first excited state $G =1^+$ plunges into the lower band, some quarks will move into it. At this moment the baryon density profile inside the cell, which was initially peaking at the center as corresponds to a well localized soliton, begins to change. The appearance of one excited quark within the cell leads to a modification of the baryon density distribution, which develops a minimum at the center of the cell and two bumps on each side. This  mechanism suggests, with the caveat expressed above, a transition to the $B=1/2$ quarkyonic phase at intermediate densities.  This suggested relation is in line with the close connection found between the discrete half-skyrmion symmetry and the continuous chiral symmetry \cite{Forkel:1989wc}.

Summarizing, the bosonized half-skyrmion phase, from the point of view of a fermionic description, is characterized by the fact that there are two types of quarks according to their spatial distribution, those which are in the $G=0^+$ band and those which are in the $G=1^+$ band.

At densities larger than $\rho_0$, the band keeps broadening and before the breakdown of solution takes place, the upper and the lower bands merge. The quarks are now free to populate also the excited states and therefore free to move everywhere in the lattice, this being the behavior of the deconfined phase as discussed in Ref.~\cite{Hahn:1987xr}.

We recall  that we have kept the dilaton field frozen at its vacuum value and have not allowed for quantum fluctuations. It has been shown, that the full implementation of  scale invariance in chiral models allows for a better description of the interaction at higher densities and temperatures, and defines an interplay between scale and chiral symmetry which leads to an orthodox description  of the restoration of chiral symmetry, i.e., the vanishing of the expectation value of the $\sigma$ field ~\cite{Bonanno:2008tt,Lee:2003eg, Park:2008zg}.

The analysis presented here is built upon a series of approximations, but we are certain that the delocalization mechanism will survive  more sophisticated analyses because it is physically very intuitive. The main result of our discussion is that inside the quarkyonic phase scenario, within the structural changes that nuclear matter could undergo before deconfinement, the half-skyrmion phase arising from bosonized QCD might take place along with the restoration of QCD symmetries. The fact that we are not considering in our approximations the effect of the long range tale of the interaction might have an influence on the detailed values of the density at which the phase transition takes place but we do not expect any qualitative change in the mechanism explored.

\section*{Acknowledgements}
We would like to thank Alessandro Drago and Byung Yoon Park for useful discussions regarding their work. Our research was supported by the INFN-MINECO agreement "Hadron structure in hot and dense medium".
V. Mantovani Sarti  and V. Vento  have been also funded by the Ministerio de Econom\'{\i}a y Competitividad 
and  EU FEDER under contract FPA2010-21750-C02-01, by Consolider Ingenio 2010
CPAN (CSD2007-00042) and by Generalitat Valenciana: Prometeo/2009/129.

%%\clearpage
\FloatBarrier
\bibliography{bibliohalf.bib}
\bibliographystyle{h-physrev3.bst}

\end{document}